\documentstyle[aps,prb,epsf,preprint]{revtex} 
\begin{document}
\newcommand{\be}{\begin{equation}}
\newcommand{\ee}{\end{equation}}
\newcommand{\bea}{\begin{eqnarray}}
\newcommand{\eea}{\end{eqnarray}}
\newcommand{\beann}{\begin{eqnarray*}}
\newcommand{\eeann}{\end{eqnarray*}}
\newcommand{\fr}{\frac}
\newcommand{\df}{\stackrel{\rm def}{=}}
\newcommand{\artanh}{\rm artanh}
\newcommand{\arcosh}{\rm arcosh}                                                         
\newcommand{\gs}{\gamma_{\perp}}
\newcommand{\gl}{\gamma_{\rm L}}
\newcommand{\vs}{v_{\perp}}
\newcommand{\vl}{v_{\,\rm L}}
\newcommand{\qs}{q_{\perp}}
\newcommand{\ql}{q_{\,\rm L}}
\newcommand{\Ms}{M_{\perp}}
\newcommand{\p}{\partial}
\newcommand{\ra}{\rangle}
\newcommand{\la}{\langle}
\newcommand{\nn}{\nonumber}
\newcommand{\my}{\mu}
\newcommand{\ny}{\nu}
\newcommand{\del}{\delta}
\newcommand{\ba}[1]{\mbox{$\begin{array}{#1}$}}
\newcommand{\ea}{\end{array}}
\newcommand{\li}{\left}
\newcommand{\re}{\right}

\draft
\title{On the Short-Time Compositional Stability of Periodic Multilayers}
\author{Martina Hentschel$^a$ $^*$, Manfred Bobeth$^a$, Gerhard Diener$^b$, 
and Wolfgang Pompe$^a$}
\address{\phantom.$^a$ Institut f\"{u}r Werkstoffwissenschaft,
Technische Universit\"{a}t Dresden, D-01062 Dresden, \\ Germany}
\address{\phantom.$^b$ Institut f\"{u}r Theoretische Physik,   
Technische Universit\"{a}t Dresden, D-01062 Dresden, \\ Germany}
\maketitle

\begin{abstract}
The short-time stability of concentration profiles in coherent periodic 
multilayers consisting of two components with large miscibility gap is 
investigated by analysing stationary solutions of the Cahn-Hilliard 
diffusion equation. 
The limits of the existence and stability of periodic concentration profiles 
are discussed as a function of the average composition for given multilayer 
period length.
The minimal average composition and the corresponding 
layer thickness below which artificially prepared layers
dissolve at elevated temperatures are calculated as a function of the
multilayer period length for a special
model of the composition dependence of the Gibbs free energy.
For period lengths exceeding a critical value,
layered structures can exist as metastable states in a certain 
region of the average composition.
The phase composition in very thin individual layers, comparable with 
the interphase boundary width, deviates from that of the 
corresponding bulk phase.
\end{abstract} 

\vspace{1cm}



\section{Introduction}              \label{sec:mot}

Modern techniques of thin film deposition permit the preparation of 
multilayers with nearly arbitrary concentration profiles. 
The knowledge of the stability of artificial multilayers at elevated 
temperatures is of great practical interest. 
With increasing atomic mobility, compositional changes occur 
where the diffusion distance is determined by the mobility and
the available time. 
In the present paper, the case of coherent periodic multilayers 
consisting of two immiscible components A and B 
is considered. 
Accordingly, competing driving forces for compositional changes in these multilayers 
are the reduction of the energy of mixing of the two components and the 
reduction of the interfacial energy of interphase boundaries.

To evaluate the thermal stability of multilayers, we investigate 
in the following the early stage of compositional changes  
perpendicular to the individual 
layers within a one-dimensional model ignoring lateral perturbations 
of the layered structure as well as boundary effects at the multilayer surface 
and the interface to the substrate. 
The neglect of boundary effects is justified as long as the characteristic 
length of atomic diffusion for the considered time scale 
is small compared to the total multilayer thickness.
The compositional evolution is studied 
within the framework of the nonlinear Cahn-Hilliard diffusion equation 
\cite{cahnhill,cahn,langer,spaepen,binder}. 
Correspondingly, the multilayer is described by a smooth concentration profile
where individual layers are separated by diffuse interphase boundaries.
The continuum description seems questionable in the limiting case
of very thin layers of a few monolayers only.
However, many predictions of the continuum approach agree qualitatively 
with those of a detailed analysis of an appropriate lattice model by Hillert \cite{hillert}. 

The composition in multilayers evolves quite differently, 
depending on the initial concentration profile. 
This is illustrated by the three examples in Fig. \ref{fig_intro}. 
The as-prepared multilayers in Fig. \ref{fig_intro} were assumed as periodic 
A/B layer stacks of pure individual layers, with the exception of case (c) 
which exhibits a small thickness perturbation.
The curves show the concentration of component B 
as a function of position. They were obtained as numerical solution 
of the Cahn-Hilliard diffusion equation.
Depending on the initial layer thicknesses,
different cases can be distinguished: 
(a) dissolution of very thin layers despite the immiscibility of the 
components (phase separation can occur only on a larger length scale), 
(b) relaxation of a periodic structure to a stationary state with smooth
concentration profile, and
(c) rapid smoothening of the profile followed by slow thinning of 
thinner layers up to their complete dissolution. 
The main driving force for the composition changes in Fig. 1
is the minimisation of  interface energy.  
This happens even at the expense of volume energy so that the minimal and 
maximal values of the concentration profile generally differ
from the concentrations of the corresponding bulk phases. 
A series of experimental findings 
\cite{naja,shen,gent,delo,krawietz,chines,thibault} 
reported recently seems to be related to these peculiarities.

During mechanical alloying, the formation of nonequilibrium supersaturated 
phases of immiscible components as e. g. Ag--Cu \cite{naja,shen} and 
Co--Cu \cite{gent,delo} was observed. 
Due to the small layer thicknesses of the lamellar structure 
which arises during ball milling, a partial intermixing of the 
two components could be energetically favourable.
Similarly, an enhanced solution of carbon in nickel layers was  
detected in Ni/C multilayers prepared by pulsed laser evaporation
with individual layer thicknesses of only a few nanometers \cite{krawietz}. 
Another surprising observation is the formation of a mixed Co--Cu phase 
during annealing of Co/Cu multilayers despite the immiscibility of 
Co and Cu \cite{chines}.
The metastable mixed phase formed obviously due to the large excess of 
interface energy in the multilayer.
A strong intermixing was also found 
during deposition of a few monolayers Ni onto Au \cite{thibault}.
In this case, besides interface energy, the large elastic energy 
of the strained Ni layer is an additional driving force for intermixing.
For a better understanding of these experimental findings, 
the present theoretical work deals with the effect of the high portion
of interface energy on the early composition evolution 
in nanoscale multilayers. 

The simulations in Fig. \ref{fig_intro} as well as numerical and analytical 
investigations by other authors \cite{langer,tsaka} reveal that concentration 
profiles evolve in characteristic stages: 
(i) the relaxation to layered quasi-stationary states or 
the complete vanishing (dissolution) of very thin layers takes place 
comparatively rapidly; 
(ii) at a longer time scale, a slow ripening process involving diffusion between
distant layers occurs, i. e. a thinning of the thinnest layers and a 
corresponding thickening of the thicker ones. 
In the following, the slow ripening process 
is referred to as long-time composition evolution, whereas the relaxation 
of an arbitrary periodic concentration profile to
a stationary solution of the Cahn-Hilliard diffusion equation is referred to as 
short-time evolution (Fig. \ref{fig_intro}b). 
Also, the rapid dissolution of layers as shown in Fig. \ref{fig_intro}a 
is considered as a particular case of the short-time evolution. 

The aim of the present work is to analyse the conditions under 
which either a rapid dissolution of thin layers occurs initially 
(Fig. \ref{fig_intro}a) or a stationary periodic concentration profile evolves 
(Fig. \ref{fig_intro}b). 
To this end, we focus on one-dimensional stationary solutions of the Cahn-Hilliard 
diffusion equation which are characterised by
the multilayer period $d$ and the average composition $\bar c$ of the multilayer. 
These two parameters are usually controlled by the 
multilayer preparation and are conserved during relaxation towards
quasi-stationary states. 
From the following analysis, a $\bar c$-$d$ diagram results
which shows the existence region as well as the
stability properties of the stationary solutions.
In the present paper, a periodic solution is called (globally) stable
if its energy is smaller than those of all other concentration profiles
with the same periodicity $d$ and average composition $\bar c$. All other
stationary solutions are metastable or unstable with respect to perturbations
conserving the periodicity and average composition. 
As outlined by Langer \cite{langer}, stationary periodic 
solutions are always unstable 
against perturbations of the periodicity (cf. also Fig. \ref{fig_intro}c). 
However, as mentioned above, these periodicity perturbations develop in 
general on a longer time scale and are not considered in this work. 
Also, small {\it lateral} composition perturbations as well as a roughening of 
interfaces between strained inidvidual layers \cite{srid} are not analysed here 
because it is expected that such perturbations develop on a longer time scale.

The paper is organised as follows. After the description of the 
model in Sect. II, stationary solutions
of the Cahn-Hilliard diffusion equation are considered in Sect. III.
Sect. IV deals with a specific model for the Gibbs free energy which
allows an analytical calculation of stationary concentration profiles.
In Sect. V, the stability of these concentration profiles is investigated.
Finally, the results are discussed and summarised.

\section{Model}                                 

The Gibbs free energy density $f(c)$ of a binary A-B system 
as a function of the {\it uniform} concentration $c$ (mole fraction) of 
component B is to exhibit two minima. 
The equilibrium concentrations, $\alpha$ and $\beta$, 
of large coexisting phase regions 
(strictly two half-spaces) are determined by the common
tangent construction (Fig. \ref{fig_fvonc}).
In the present work, the case of multilayers with planar interfaces
is considered. 
Following Cahn and Hilliard \cite{cahnhill},  
the free energy per unit area of a system with concentration $c(x)$ 
varying in one dimension is described by
\begin{equation}  \label{f_cht}
        F[c] = \int dx \left[ f(c) + \kappa \left( \fr{d c}{d x}\right)^2
                \right]  \:. 
\end{equation}
The second term on the right-hand-side of (\ref{f_cht}) represents the energy 
contribution due to a concentration gradient where $\kappa$ is the  
gradient energy coefficient.

The interdiffusion flux in the system is given by 
$j(x) = - \tilde M \, \partial ( \delta F / \delta c) / \partial x$ 
where $\tilde M$ is the atomic mobility. 
Together with the continuity equation 
$\partial c / \partial t  + \Omega \, \partial j / \partial x = 0$,
the following nonlinear diffusion equation results \cite{cahn,spaepen,binder}
\be     \label{ch_eq}
        \fr{\p c}{\p t}  = 
                M \; \fr{\p^2 }{\p x^2} \left( \fr{\del F[c]}{\del c(x)} \right) =
                M \; \fr{\p ^2}{\p x^2} \left[f'(c) - 2 \kappa \fr{\p ^2c}{\p x^2}\right] \:
\ee
where $M \equiv \Omega \,  \tilde M$ with $\Omega$ the atomic volume. For simplicity,
the atomic volume of the two components has been assumed to be equal and the
composition dependence of $M$ has been omitted.
Starting from any initial concentration profile, the further
evolution can be calculated numerically from (\ref{ch_eq}) (see 
e. g. Fig. 1). 
However, in view of the great practical importance of 
quasi-stationary concentration profiles, we will
analyse the stationary solutions of the Cahn-Hilliard diffusion equation 
(\ref{ch_eq}) more systematically in the following.

\section{Stationary solutions}                          

Equilibrium concentration profiles are determined by the
extrema of the free energy under the constraint of particle conservation.
This leads to the variational problem
\be \label{statlos}
        \fr{\del}{\del c(x)} \left( F[c]-\mu \int dx \, c(x) \right)  = 0   \;
\ee
with the result
\be \label{el_eq}
        f' (c)  - 2 \kappa \fr{d^2 c}{d x^2} - \mu = 0 
\ee
(the prime denotes the derivative with respect to $c$).
Comparison of (\ref{el_eq}) and (\ref{ch_eq}) reveals
that solutions of (\ref{el_eq}) are also stationary 
solutions of the diffusion equation (\ref{ch_eq}). 
The Lagrange multiplier $\mu$ is identified as interdiffusion potential. 
When $\mu = \delta F / \delta c$ is uniform, the particle 
flux vanishes.  
Integration of (\ref{el_eq}) leads to 
\be     \label{firint1}
        \kappa \left(\fr{d c}{d x}\right)^2 = f(c) - \mu c + K \equiv D(c)
\ee
where $K$ is an integration constant. The last equality in (\ref{firint1})
defines the function $D(c)$ used in the following. 
In general, the physically relevant solutions of (\ref{firint1}) are  
periodic concentration 
profiles $c(x)$ oscillating between a minimal value $a$ and a maximal 
value $b$ (cf. Figs. 2 and 3). 
The extrema $a$ and $b$ are related to the parameters $\mu$ and $K$ by 
the conditions $D(a) = D(b) = 0$ which lead to  
\be     \label{muk}
        \mu = \frac{f(b) - f(a)}{b - a}, \qquad K = 
                \frac{f(b) \, a - f(a) \, b}{b - a} \: .
\ee
Further integration of (\ref{firint1}) yields the inverse function of the 
concentration profile
\be     \label{xvc1}
        x(c)=\int_{c_0}^c dc \; I(c)
\ee
with $I(c) = \sqrt{\kappa / D(c) } = dx/dc$. 
The integration bounds have been chosen in such a way that the origin 
of the $x$-coordinate is located at an interphase boundary 
defined by $c = c_0 \equiv (a+b)/2$.
Thus, equation (\ref{xvc1}) represents the concentration 
profile in half a period from 
$c = a$ at $x = - d_a/2$ to $c = b$ at $x = d_b/2$ (Fig. \ref{fig_cvonx}).
The individual layer thicknesses $d_a$ and $d_b$ of the two phase regions,
briefly called phase 'a' and phase 'b', are given by  
\be     \label{dadb}
        d_a = 2 \, \int_a^{c_0} dc \,I(c)  \:, \qquad
        d_b = 2 \, \int_{c_0}^b dc \,I(c)  \:.
\ee
Similarly, the multilayer period length $d = d_a + d_b$ and the
mean composition $\bar c$ are given by
\be     \label{dcm}
        d  = 2 \, \int_a^b dc \,I(c)  \:, \qquad
        \bar c = \frac{2}{d} \int_{- d_a/2}^{d_b/2} dx \, c(x) = 
                 \frac{2}{d} \int_a^b dc \,I(c) \, c \:.
\ee

For given concentrations $a$ and $b$, the concentration profile $c(x)$ 
can be calculated directly from (\ref{xvc1}).
However, from the experimental point of view, 
$a$ and $b$ are not known {\it a priori}. Usually, components
A and B are deposited consecutively with fixed individual layer thicknesses. 
During the early annealing stage, 
a smooth concentration profile develops 
which is similar to the stationary profiles derived here. 
The arising concentrations $a$ and $b$ are determined by equations
(\ref{dcm}) where the mean composition $\bar c$ and period length $d$
are given.
The solution of equations (\ref{dcm}) for $a$ and $b$ is, however, not 
always unique.
If there is more than one solution, one has to compare the free energies 
of different solutions in order to find that with the lowest one. 
From (\ref{f_cht}) and (\ref{firint1}),
the free energy of one multilayer period results as 
\be     \label{energy}
        F_{p} = 4 \sqrt{\kappa} \int_a^b dc \, \sqrt{D(c)} \;+ \; 
                  \left[ \frac{b - \bar c}{b - a} f(a) + 
                        \frac{\bar c - a}{b - a} f(b) \right ] \; d \; .
\ee

An important intrinsic length of the present problem is the 
width of the interphase boundary $\xi$ defined by
\be     \label{xi}
        \xi = (b - a) \; \frac{dx(c)}{dc} |_{c=c_0}  
            = (b - a) \; \sqrt{\frac{\kappa}{D(c_0)}} \; .
\ee
The second equality follows from (\ref{firint1}). 
In the limiting case of spatially extended phases  
($d_a, d_b \gg \xi$; i. e. $a \rightarrow \alpha$, $b \rightarrow \beta$), 
the interface width becomes 
\be     \label{xi0}
        \xi = (\beta - \alpha) \; \sqrt{\kappa/f_0} \; 
            \equiv (\beta - \alpha) \, l_0 \:
\ee
with $f_0 \equiv D((\alpha + \beta)/2)$.
$f_0$ characterises the height of the free energy wall of the $f(c)$ 
curve referred to the common tangent (cf. Fig. \ref{fig_fvonc}).
The last equality in (\ref{xi0}) defines the length unit $l_0$ used 
in the following.

For very thin individual layers, comparable with the width of the interphase 
boundary, the concentrations in the middle of the layers, $a$ and $b$, 
differ from those of the corresponding bulk phases $\alpha$ and $\beta$ 
($\alpha < a <  b < \beta$, Fig. \ref{fig_cvonx}) because 
the common tangent construction does not apply to thin layers.
The concentrations $a$ and $b$ define a secant with 
the $f(c)$ curve as shown in Fig. \ref{fig_fvonc}.
An estimate of the difference between the concentrations 
$\beta$ and  $b$ is given by 
\be     \label{dick_dien}
        \fr{\beta - b}{\beta -\alpha}  =  \rho_b \, \exp( - d_b / 2 \xi_b) 
\ee
(see Appendix A), where $\xi_b \equiv \sqrt{2 \kappa/f''(\beta)}$ 
and $\rho_b$ is a numerical factor of the order of unity.
An analogous formula applies to the difference $a - \alpha$.
In the limiting case $d_b \gg \xi_b$, a factor of $\rho_b = 2$ 
has been calculated for the special composition dependence of the 
Gibbs free energy considered in the following section. 
The estimate (\ref{dick_dien}) clearly reveals that 
concentrations $\beta$ and $b$ differ significantly when the layer thickness
$d_b$ approaches the characteristic length $\xi_b$.

\section{Special case: $c^4$--model}                   

To simplify the calculation of the concentration profile (\ref{xvc1}), 
we consider the case where the
free energy as a function of concentration can be represented as a
polynomial of the fourth power
\be \label{fvonc_c4}
        f(c) = A (c-\alpha)^2 (c-\beta)^2 + B (c-\alpha) + C \:.
\ee
The parameters $B$ and $C$ turn out to be unimportant for the composition evolution.
The characteristic energy unit $f_0$ results as $f_0 = A \, (\beta - \alpha)^4/16$. 
The parameters $\alpha$ and $\beta$ in (\ref{fvonc_c4}) coincide with the equilibrium
concentrations of spatially extended phases corresponding to 
the common tangent construction (Fig. \ref{fig_fvonc}). 
The characteristic length $\xi_b$ in (\ref{dick_dien}) is obtained as 
$\xi_b = (\beta - \alpha) \, l_0 / 4$. 

For the case (\ref{fvonc_c4}), briefly called '$c^4$-model', 
the inverse concentration profile (\ref{xvc1}),
the period length, and the mean composition (\ref{dcm}) can be expressed 
by elliptic integrals
\be     \label{c4konzprof}
        x(c) = \omega \, \sqrt{\fr{\kappa}{A}}\:\, 
                 [F_e(\phi(c),m) - F_e(\phi(c_0),m)] \;,
\ee
\be
        d = 2 \, \omega  \, \sqrt{\fr{\kappa}{A}} \: K(m) \;, 
\ee
\be 
        \bar{c} = a + \fr{2}{\omega} \:Z(\phi_Z,m) \;
\ee
with $\omega=2/\sqrt{(b_1 -a)(b-a_1)}$, $m=(b-a)(b_1-a_1)/((b-a_1)(b_1-a))$,
$\phi_Z = \arcsin\sqrt{(b_1-a)/(b_1-a_1)}$, and  
$\phi(c) = \arcsin{\sqrt{(b-a_1)(c-a)/((b-a)(c-a_1))}}$.
$K(m)$ and $F_e(\phi,m)$ are the complete and incomplete elliptic integrals of
first kind, and $Z(\phi,m)$ is the Jacobi zeta function \cite{abram}. 
$a$ and $b$ are the concentrations in the middle
of the individual layers, and $a_1$ and $b_1$ are the further 
two intersections of the $f(c)$-curve with the secant shown in 
Fig. \ref{fig_fvonc}.
$a_1 < a < b < b_1$ are also
the four zeros of the function $D(c)$ defined in (\ref{firint1}).
Choosing $a$ and $b$, the other two zeros are given by  
\be  \label{eq18}
        a_{1} = \alpha + \beta -c_0 - w, \qquad
        b_{1} = \alpha + \beta -c_0 + w
\ee
with 
$w \equiv \sqrt{ 2 \, [ c_0 (\alpha + \beta) - \alpha \beta] + ab - 3 c_0^2}$
and $c_0 \equiv (a+b)/2$.  
Analytical expressions for the stationary solutions of the
Cahn-Hilliard diffusion equation in the case of the $c^4$-model have been 
given previously in terms of Jacobian Elliptic functions by Tsakalos 
\cite{tsaka} for the symmetric case $d_a = d_b$ and by 
Novick-Cohen and Segel for the general case \cite{novick-cohen}. 

Fig. \ref{fig_fvsa} shows a series of interesting quantities as a function of 
the mean concentration $\bar c$ for fixed period length $d$, 
calculated within the $c^4$-model.
In a certain $\bar c$--$d$ region, two stationary periodic solutions 
have been found. 
The corresponding branches in Fig. \ref{fig_fvsa} are denoted
by \#1 and \#2, respectively.  
The free energy $F_{p}$ of the periodic concentration profile 
with smaller concentration variation $b-a$ (solution 2) 
is higher than that of solution 1 (Fig. \ref{fig_fvsa}a).
For small values of $\bar c$, 
the free energy of the homogeneous concentration, 
$F_h = f(\bar c) \; d$, is lower than that of the periodic 
solution 1. The values of the concentrations $a$ and $b$ are shown
in Fig. \ref{fig_fvsa}b. A striking feature of these plots
is the existence of a minimal value $\bar c_a^M$ of the mean concentration.
At the minimum $\bar c_a^M$, solutions 1 and 2 merge.
The corresponding layer thickness of phase 'b' is denoted by 
$d_b^{M}$  (Fig. \ref{fig_fvsa}c). It correlates roughly
with the minimal value of $d_b$ in this case. 
As discussed in the next section,
solution 2 does not appear for multilayer period lengths $d$ below
a critical value.

\section{Stability of stationary solutions}             

In the following, let us consider the stability of periodic concentration 
profiles $c(x)$ which are obtained as solutions of equation (\ref{firint1}). 
These concentration profiles are also stationary solutions of the 
Cahn-Hilliard diffusion equation  (\ref{ch_eq}). 
They are stable against {\it any} infinitesimal perturbation $\del c(x)$ 
if the second variation of the free energy  
\be     \label{zw_var}
        \del^2 F =  \int  dx  \left[ f'' (c) (\del c)^2 + 
                             2 \kappa \left( \fr{d \del c}{dx}  \right)^2 
                      \right]     \: 
\ee
is positive definite. 

At first, the stability of the homogeneous concentration $c(x) = \bar{c}$ 
is considered. 
Without any restriction, the perturbation $\del c$ can be represented
as a Fourier series. 
From (\ref{zw_var}), it is evident that the stabilising influence of the 
gradient term (second term on the right-hand-side of (\ref{zw_var})) is
stronger the shorter the wavelength of the perturbation is.
Considering a periodic perturbation with period $d$, i. e. 
$\del c \propto \sin(2 \pi x/d)$, we find stability $\del^2 F \ge 0$
for period lengths
\be  \label{dmin}
       d < d^S(\bar{c}) = 2 \pi \sqrt{\fr{2 \kappa}{- f''(\bar c)}}  \:.
\ee
$d^S(\bar c)$ is equal to the smallest wavelength of spinodal decomposition 
obtained from a stability analysis of diffusion equation (\ref{ch_eq}) 
\cite{spaepen,binder}.
The inverse function of $d^S(\bar c)$ exhibits two branches which
are denoted by $\bar c_a^S(d)$ and $\bar c_b^S(d)$ 
(Fig. \ref{fig_stabdia}). 
For the $c^4$-model, one obtains 
\be  \label{cmvsd}
         \bar{c}_{b/a}^S(d) = \fr{\alpha + \beta}{2} \, \pm \,
                \fr{\beta - \alpha}{2 \sqrt{3}} \,  
                \left[ 1 - \fr{\pi^2}{2} (\beta - \alpha)^2 
                \left (\fr{l_0}{d} \right )^2 \right]^{1/2}    \:.
\ee
Within the interval $\bar c_a^S < \bar c < \bar c_b^S$, the homogeneous solution is 
unstable and spinodal decomposition takes place.
According to (\ref{dmin}), the maximum of 
$-f''(\bar c)$ yields a minimal period length $d_{min}$
below which the homogeneous solution is stable for all values of $\bar c$.
For the $c^4$-model,  
$d_{min} = \pi (\beta-\alpha) \,l_0/\sqrt{2}$ follows.

Since within the region $\bar c_a^S < \bar c < \bar c_b^S$ the homogeneous 
solution is unstable, there must be a stable periodic solution.
The investigation of the existence region and stability of periodic solutions 
is more complicated than that of the homogeneous one.
In the following, the results of a numerical calculation of the free 
energies (10) belonging to the solutions (\ref{c4konzprof}) of the $c^4$-model 
are summarised (Fig. \ref{fig_stabdia}). 
It is expected that the qualitative features are the same for 
other double-well potentials $f(c)$. Similar results have been obtained
by Hillert \cite{hillert} for a lattice model.  

The behaviour of periodic solutions changes qualitatively 
in dependence on the multilayer period length $d$. 
The stability of concentration profiles can be discussed in a 
convenient way by including non-stationary states with nonequilibrium 
amplitudes $b-a$ in the consideration. 
Although the parameter $b-a$ does not give a
complete characterisation of non-stationary states, it is an appropriate 
quantity to illustrate the stability behaviour of periodic concentration 
profiles. 
In Fig. \ref{fig_fvsba}, the free energy $F$ of concentration profiles 
is sketched as a function of their amplitude for different mean 
compositions and for two multilayer period lengths.
Extrema of $F$ correspond to stationary solutions. 
In Fig. \ref{fig_fvsba}, these extrema are marked by filled and open 
circles denoting stable and unstable stationary solutions, 
respectively.

For period lengths above $d_{min}$, but below a certain critical 
value $d_c$, the situation is re\-presented by the curves 1 and 2 in
Fig. \ref{fig_fvsba}a:
(i) Within the region $\bar c_a^S < \bar c < \bar c_b^S$,
there is a stable periodic solution corresponding to the minimum
of $F$ (curve 1). The homogeneous solution ($b-a=0$) belongs to
a maximum of $F$ (more precisely a saddle point) and is therefore unstable.
(ii) Outside the region $\bar c_a^S < \bar c < \bar c_b^S$, there
is no stationary  periodic solution (curve 2). The minimum of the
free energy $F$ is given by the homogeneous solution 
which, therefore, is stable. 
At the concentrations $\bar c = \bar c_{a,b}^S$, 
a continuous transition between the homogeneous 
and the periodic stationary solution occurs. 
This transition will be referred to as second order 'phase transition' 
in the following despite the fact that the resulting 'phases'
are only stable in the sense discussed in the Introduction. 

For larger period lengths $d > d_c$, the behaviour is more complex 
as illustrated in Fig. \ref{fig_fvsba}b. 
Within the region $\bar c_a^S < \bar c < \bar c_b^S$, the
situation is the same as for $d < d_c$ (curve 1). 
Outside certain limiting values $\bar c_{a,b}^M$ of the mean concentration,
there is no stationary periodic solution (curve 4). 
However, within the regions 
$\bar c_a^M < \bar c < \bar c_a^S$ and $\bar c_b^S < \bar c < \bar c_b^M$,
two stationary periodic solutions exist (curves 2 and 3;
cf. also Fig. \ref{fig_fvsa}). 
One solution (solution 1) corresponds to a minimum of the
free energy $F$ and the other one (solution 2) to a maximum.
Thus, solution 2 is unstable.
Choosing solution 2 as initial condition for the concentration
profile, the numerical solution of the diffusion equation (\ref{ch_eq})
reveals a rapid change of the profile. 
Depending on the initial fluctuations, it evolves either into the
periodic solution 1 or into the homogeneous solution.
The latter one corresponds also to a minimum of $F$. 
Whether the periodic solution 1 or the homogeneous one
is more stable depends on the corresponding 
values of the free energy $F_p^{(1)}$ and $F_h=f(\bar c)\,d$.
At a certain concentration $\bar c = \bar c_{a,b}^T(d)$, for which
$F_p^{(1)} = F_h$, a first order 'phase transition' between
the periodic and the homogeneous states takes place
(cf. Fig. \ref{fig_stabdia}).
Between $\bar c^S(d)$ and $\bar c^T(d)$
the homogeneous solution is metastable ($F_{p}^{(1)} < F_h$,
curve 2 in Fig. \ref{fig_fvsba}b), 
whereas between $\bar c^T(d)$ and $\bar c^M(d)$ 
the periodic solution is metastable (curve 3 in Fig. \ref{fig_fvsba}b).

At $\bar c = \bar c^S(d)$, solution 2 disappears by merging 
into the homogeneous solution. 
With increasing distance of $\bar c$ from $\bar c^S$, the difference 
in the free energies of solutions 1 and 2 decreases and at the 
concentrations $\bar c^M$ they coincide (cf. also Fig. \ref{fig_fvsa}). 
This implies that the second variation of the
free energy functional vanishes (see Appendix B) and, consequently,
the periodic solutions become marginally stable at $\bar c^M$. 
Outside the region $\bar c_a^M < \bar c < \bar c_b^M$, 
no inhomogeneous stationary solution exists. 
Thus, $\bar c_a^M(d)$ and $\bar c_b^M(d)$ represent the
minimal and maximal mean concentrations for the existence of metastable
or globally stable periodic structures with given multilayer period length.

The critical value of the multilayer period $d_c$, which
separates the regions of first and second order 
'phase transitions' in Fig.  \ref{fig_stabdia}, can be derived 
analytically by means of a third-order perturbation analysis. 
The corresponding critical concentrations $\bar c_c = c^S(d_c)$ are 
determined by the equation $(f''')^2 + \, 3 \, f'' \, f'''' = 0$,
where $f''$, $f'''$ and $f''''$ denote the second to fourth derivative
of $f(c)$ at $c = \bar c_c$. 
According to (\ref{dmin}), the critical period length is then given by
$d_c = 2 \pi \sqrt{ \, 2 \kappa / (- f''(\bar c_c))}$.
For the $c^4$-model, one obtains 
$\bar c_c = [\alpha + \beta \,\pm \,(\beta - \alpha)/\sqrt{5}]/2$
and $d_c = \sqrt{5/2} \; d_{min}$.

Let us recall that the above picture on the stability of periodic solutions 
was obtained for the special $c^4$-dependence of the Gibbs free energy. 
Further analysis of other dependencies $f(c)$ is desirable to confirm the present
stability diagram qualitatively and to study quantitative changes.
Numerical simulations of the composition evolution in the present work
were performed by means of a finite difference method described in 
ref.\cite{copetti}. 
This method is based on a semi-implicit finite difference 
scheme coupled with a fast Fourier transformation.
For numerical details, the reader is referred to 
Copetti and Elliott \cite{copetti}.
The spatial grid spacing in our calculations was less than 10\%
of the interphase boundary width of stationary states
and for the time integration an adaptive step size control was applied.
For example, 1024 grid points were used for the calculations in Fig. 1.

\section{Discussion}                                    

Multilayers for practical applications are usually deposited as
layer stacks of nearly pure layers of components A and B 
with thicknesses $d_A$ and $d_B$.
During annealing, a considerable interdiffusion between individual layers
can occur depending on the values of the equilibrium bulk concentrations
$\alpha$ and $\beta$,  which change with temperature.
Consider for definiteness the case of thin layers of component B 
between thick layers of component A ($d_B < d_A$).
For arbitrary multilayer period lengths, the B-layers dissolve 
at elevated temperatures if the mean composition $\bar c = d_B/d$
(supposing $\Omega_A = \Omega_B$)
is smaller than $\alpha$, where $\alpha$ increases with
increasing temperature. 
However, layer dissolution can
occur also for $\bar c > \alpha$, if the multilayer period is 
small enough. 

The composition evolution in multilayers is controlled 
by the competition between volume free energy and interface energy.
For mean compositions outside the region
$\bar c_a^M(d)  < \bar c < \bar c_b^M(d)$, 
no stationary periodic solutions of the Cahn-Hilliard diffusion 
equation have been found.
This implies that individual layers of such multilayers dissolve 
in the early stage of annealing because
the free energy gain by phase separation is too small to 
overcompensate the interface energy (Fig. 1a). 
The thickness of the 'b'-layers $d_b^M \equiv d_b(\bar c = \bar c_a^M)$,
which corresponds to the minimal mean concentration $\bar c_a^M(d)$, 
is shown in Fig. \ref{fig_lbmin} as a function of the period length $d$.
For large period lengths ($d > 7 \, l_0$ for the example in 
Figs. \ref{fig_stabdia} and \ref{fig_lbmin}), $d_b^M(d)$ represents 
the minimal layer thickness of phase 'b' of stable concentration profiles. 
For smaller $d$ (but larger than $d_c$), the minimal thickness 
$d_b$ appeared in the present case at a mean concentration slightly 
larger than $\bar c_a^M$ (cf. Fig. \ref{fig_fvsa}c). 

Despite the large change of the multilayer period length in Fig. 
\ref{fig_lbmin}, the characteristic layer thickness $d_b^M(d)$ changes only 
slightly and is always of the order of the characteristic length 
$l_0 = \sqrt{\kappa / f_0}$. 
The gradient energy coefficient $\kappa$ can be estimated within the framework 
of the regular solution model 
\cite{cahnhill,spaepen} as $\kappa \approx \Delta U /r_0$ 
where $r_0$ is the interatomic distance and 
$\Delta U$ is the energy of mixing (per atom) of an equiatomic solution 
($\bar c = 0.5$);
typically, $\kappa = 10^{-11}$ to $10^{-10}$ J/m.
The value of $l_0$ considerably exceeds the interatomic distance 
when $f_0 \ll \Delta U / r_0^3$, where  
$\Delta U / r_0^3$ is typically in the range of $10^{8}$ to $10^{9}$ J/m$^3$.
Choosing for example  $\kappa = 3 \cdot 10^{-11}$ J/m and 
$f_0 = 10^{8}$ J/m$^3$, one obtains $l_0 \approx 0.5$ nm
which is about twice the interatomic distance.

The gradient energy coefficient changes only very
weakly with temperature, whereas the parameter $f_0$ (cf. Fig. 2)
decreases with increasing temperature, approaching zero at the critical 
phase separation temperature $T_c$. 
As a consequence, the length $l_0$ and correspondingly 
the thickness $d_b^M$, characterising the onset of layer 
dissolution,  
diverge as $|T-T_c|^{-1/2}$ at the critical temperature. 
Comparatively small critical temperatures in 
technologically important regions of a few $100^\circ$C are found,
for example, for systems with large lattice mismatch because mechanical 
stresses in coherent layers cause a considerable 
lowering of the critical temperature compared to incoherent phases
\cite{cahn}. 

The layer dissolution in multilayers with very small period length, 
which is driven by the reduction of interface energy, leads 
to the formation of supersaturated phases.
Subsequent phase separation occurs with a larger period length 
on a longer time scale and can be kine\-tically hindered by
rapid quenching. 
As mentioned in the introduction, such a situation could be present 
during mechanical alloying \cite{naja,shen,gent,delo}, 
when a nanoscale lamellar structure develops in the course of ball-milling, 
or in the case of multilayer deposition \cite{krawietz,thibault} with 
subsequent short-time annealing at moderate temperatures \cite{chines}. 
The nonequilibrium phase formation observed in those experiments 
could be related to the layer dissolution discussed here.
Based on an estimate of the chemical energy of coherent phase boundaries,
such a mechanism has been suggested by Gente et al. \cite{gent} 
to explain the observed solid solution formation of immiscible elements 
due to long-time ball-milling.

\section{Summary}                       

Stationary solutions of the one-dimensional Cahn-Hilliard diffusion 
equation have been analysed in dependence on the mean composition 
$\bar c$ of a binary system and the multilayer period length $d$.
A $\bar c$-$d$ diagram has been established
showing the regions of existence, metastability and global stability of 
stationary periodic concentration profiles, as well as of the homogeneous 
concentration (Fig. \ref{fig_stabdia}).
The diagram was derived under the constraint of fixed period length 
of the concentration profile. 
Actually, periodic solutions are unstable against thickness fluctuations 
\cite{langer} which leads to layer thickness coarsening
in the course of annealing. 
However, this happens on a longer time scale than the relaxation
to quasi-stationary concentration profiles as long as the individual layer 
thicknesses are significantly larger than the interphase boundary width. 

The present analysis revealed that very thin individual layers 
(typically a few monolayers for small mutual solubility) 
dissolve during annealing if
the mean composition of artificial multilayers is lower than 
a critical value $\bar c_a^M(d)$. 
The individual layer thickness $d_b^M$ corresponding to the minimal value 
$\bar c_a^M(d)$ increases slightly with increasing multilayer period length.
The layer dissolution is driven by two mechanisms: 
(i) interdiffusion between pure individual layers to establish 
the equilibrium phase concentrations and (ii) reduction of 
interface energy in the case of very thin layers of the order 
of the interphase boundary width. 
In a certain $\bar c$-$d$ region, the layered structure can exist as
metastable state although the free energy of the homogeneous concentration
is lower.
According to the Cahn-Hilliard theory,
the equilibrium composition in thin layers differs from that of the 
corresponding bulk phase if the layer thickness becomes comparable
with the interphase boundary width \cite{langer,tsaka,novick-cohen}.

Consideration of the present conclusions in the design of 
layered structures could help to improve their thermal stability or,
on the other hand, to prepare new metastable phases by controlled layer 
dissolution. 
Without changing the qualitative predictions,
mechanical stresses in coherent layers due to lattice 
mismatch can easily be
included in the present one-dimensional analysis
by a modification of the free energy $f(c)$ (see e. g. ref. \cite{cahn}).
 
The evaluation of the long-time stability of multilayers requires an
additional analysis of the evolution of lateral composition perturbations 
including  the effect of stresses in individual layers. 
The roughening instability of interfaces between strained layers has been 
investigated for example in ref.\cite{srid}. 
A comprehensive analysis of the composition evolution under the influence of 
stresses owing to lattice mismatch or to the presence of dislocations 
has been given recently in a series of papers by 
L\'eonard and Desai \cite{leonard}.

\acknowledgements{
This work was supported by the Deutsche Forschungsgemeinschaft, 
Sonderforschungsbereich 422.
}

\newpage

\section{Appendix A}                                    

To derive (\ref{dick_dien}),  equation (\ref{el_eq}) is linearised
with respect to the concentration difference $\delta c_b(x) = b - c(x)$. 
With the approximation  $f''(b) \approx f''(\beta)$, one obtains
\be     \label{exp2}
        \left ( \fr{d^2 }{d x^2} - 
                \fr{1}{\xi_b^2}  \right ) \,\delta c_b(x) = 
                \fr{\mu - f'(b)}{2\kappa} \;
\ee
($\xi_b^2 \equiv 2 \kappa/f''(\beta)$).
Choosing the middle of the 'b'-layer as origin of the $x$-coordinate
and requiring $\delta c_b(0) = \delta c_b'(0) = 0$, 
the solution of (\ref{exp2}) is obtained as 
\be  \label{yvx}
        \delta c_b(x)= \fr{\mu - f'(b)}{f''(\beta)}
                \left (\cosh \fr{x}{\xi_b} - 1 \right)\;.
\ee
Although this result was derived for small values of $\delta c_b$, it is 
extrapolated to larger values in order to get an estimate for the 
layer thickness $d_b$ defined by $c( \pm d_b/2 ) = (a+b)/2$.
Assuming further $d_b \gg \xi_b$ and approximating $b-a$ by $\beta - \alpha$,
one finds 
\be  \label{db}
        \delta c_b(\pm d_b/2) = \fr{b - a}{2}  \approx 
                \fr{(\beta - \alpha)}{2} =
                \fr{1}{\rho_b} \; \fr{\mu - f'(b)}{2 \, f''(\beta)} \exp(d_b / 2 \xi_b). 
\ee
The correction factor $\rho_b$  is introduced in (\ref{db})
in order to account for the error caused by the linearisation of 
(\ref{exp2}) within the interface. 
The last equation can be rewritten using the expansion
$f'(b) \approx f'(\beta) + f''(\beta) (b - \beta)$ and the fact 
that $\mu - f'(\beta)$ is of the order of
${\cal O}((\beta - b)^2,\: (a - \alpha)^2)$.
Neglecting these higher order terms and 
inserting $\mu - f'(b) \approx  f''(\beta) (\beta - b)$ into (\ref{db}),
one obtains equation (\ref{dick_dien}).

\section{Appendix B}                                    

In the following, the stability of two stationary solutions
for the same $d$ and $\bar{c}$,
$c_1(x)$ and  $c_2(x)$ differing by  
$\delta c(x) = c_2(x) - c_1(x)$, is analysed for 
$\delta c \rightarrow 0$. The two solutions fulfil
equation (\ref{el_eq}) with the corresponding Lagrange multipliers
$\mu_1$ and $\mu_2$. Expansion of (\ref{el_eq}) for 
$c_2 = c_1 + \delta c$ with respect to $\delta c$ and  
$\delta \mu = \mu_2 - \mu_1$ yields the following equation for $\delta c$:
$f''(c_1) \, \delta c - 2\kappa\,d^2 \delta c/d x^2 =  \delta \mu$.
Using the estimate 
$\delta \mu \sim {\cal O}((\delta c)^2)$,
one obtains to first order in $\delta c$
\be     \label{dcnull} 
        \left [ f''(c_1) - 2\kappa\fr{d^2}{d x^2} \right] \;\delta c(x) = 0 \:. 
\ee 
This equation is equivalent to marginal stability of $c_1(x)$.
Indeed, the second variation  (\ref{zw_var}) of $F[c]$ can be transformed by
partial integration into 
\be     \label{op_eq}
        \delta^2 F = \int dx \: \delta c \left[f''(c) - 
                2 \kappa \fr{d^2}{dx^2} \right]  \delta c  \:. 
\ee
In deriving (\ref{op_eq}), the periodicity of solutions $c_1(x)$ and $c_2(x)$,
and consequently of $\delta c(x)$, was used.
Comparison of (\ref{op_eq}) and (\ref{dcnull}) leads to $\delta^2 F = 0$. 
In summary, if two solutions merge, marginal stability results.
This happens at the boundaries $\bar c^S$ as well as $\bar c^M$ 
(Fig. \ref{fig_stabdia}).

\phantom.$^*$ Present address: Max-Planck-Institut f\"ur Physik komplexer
Systeme, N\"othnitzer Str. 38, D-01187 Dresden, Germany

\newpage

\begin{figure}

\caption{  \label{fig_intro} 
  Early composition evolution of 
  stacks of initially pure layers ($t=0$, dashed lines):  
  (a) rapid dissolution of individual layers 
      ($t/t_0 =$ 0.004, 0.02, 0.1),
  (b) relaxation to quasi-stationary profile 
      ($t/t_0 =$  1, 1000; the two curves coincide), and
  (c) fast relaxation to quasi-stationary profile with subsequent 
      slow thinning and rapid dissolution of the initially thinner layer
      ($t/t_0 =$ 0.1, 171.3, 171.5; $t_0 \equiv \kappa/ M f_0^2$, 
      parameters see Sect. II and III).
}
\vspace{1cm}

\caption{  \label{fig_fvonc} 
  Gibbs free energy density as a function of concentration of component B
  together with the common tangent and a secant corresponding to 
  infinite and finite period length $d$ of stationary concentration profiles,
  respectively. The secant intersects the $f(c)$ curve at the 
  concentrations $a_1 < a < b < b_1$.
}
\vspace{1cm}

\caption{  \label{fig_cvonx} 
  Stationary concentration profile illustrating the 
  meaning of concentrations $a$ and $b$ and the
  layer thicknesses $d_a$ and $d_b$ ($\alpha = 0.1$, $\beta = 0.9$). 
  The dashed line represents a second stationary profile 
  for the same parameters $d$ and $\bar c$,
  which is, however, unstable (cf. Sect. V).
}
\vspace{1cm}

\caption{  \label{fig_fvsa} 
   (a) Free energies of periodic concentration profiles $F_{p}$ 
   (full line) and of the corresponding homogeneous concentration 
   $F_{h}$ (dashed), (b) concentrations $a$ (full) and $b$ (dashed), as well as
   (c) layer thickness of phase 'b' as a function of the 
   mean concentration $\bar c$
   ($d = 6 \,l_0$, $\alpha =$ 0.01, $\beta =$ 0.99).
   The intersection of $F_{p}(\bar c)$ and $F_{h}(\bar c)$
   determines the phase boundary $\bar c^T$ in Fig. \ref{fig_stabdia}.
   The thickness $d_b$ at the minimal value 
   $\bar c_a^M$ defines $d_b^{M}$ in Fig. \ref{fig_lbmin}.
} 
\vspace{1cm}

\caption{  \label{fig_stabdia} 
   Phase diagram showing the phase boundary $\bar c^T(d)$ 
   between homogeneous and periodic solutions as well as the
   limits of metastability of the periodic and homogeneous 
   solutions $\bar c^{M}(d)$ and $\bar c^{S}(d)$, respectively 
   ($\alpha =$ 0.01, $\beta =$ 0.99). 
   The vertical dashed line at $d = d_c$
   separates the regions of first and second order transition at
   the concentrations $\bar c^{S}$ (2-PT) and $\bar c^{T}$ (1-PT),
   respectively.
} 
\vspace{1cm}

\caption{  \label{fig_fvsba} 
   Schematic of the dependence of the free energy on the amplitude of
   the concentration profile for two period lengths (a: $d<d_c$, 
   b: $d>d_c$). The origin of the amplitude-axis corresponds to the 
   homogeneous concentration $b-a=0$ (for further explanations see text).
} 
\vspace{1cm}

\caption{  \label{fig_lbmin} 
  Layer thickness $d_b^{M} = d_b(\bar c_a^M)$ as a function of the 
  period length $d$; in the unstable region individual layers dissolve  
  ($\alpha =$ 0.01, $\beta =$ 0.99).
}
\vspace{1cm}

\end{figure}


\newpage

\vspace*{-1cm}
\epsfxsize=15cm
\epsffile{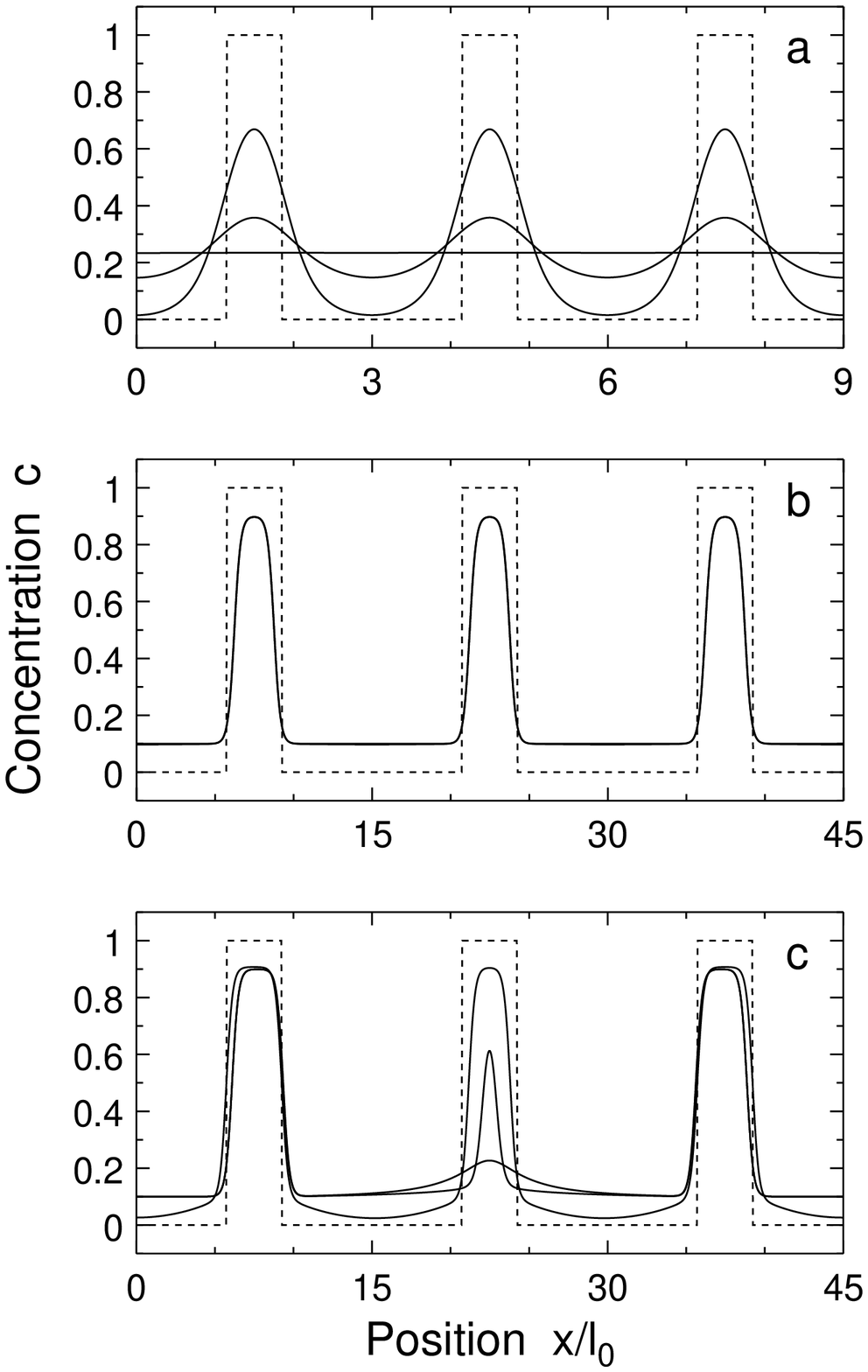}
\vspace{1cm} \hspace*{14cm} {\bf Fig. 1 }  
\newpage

\epsfxsize=16cm
\epsffile{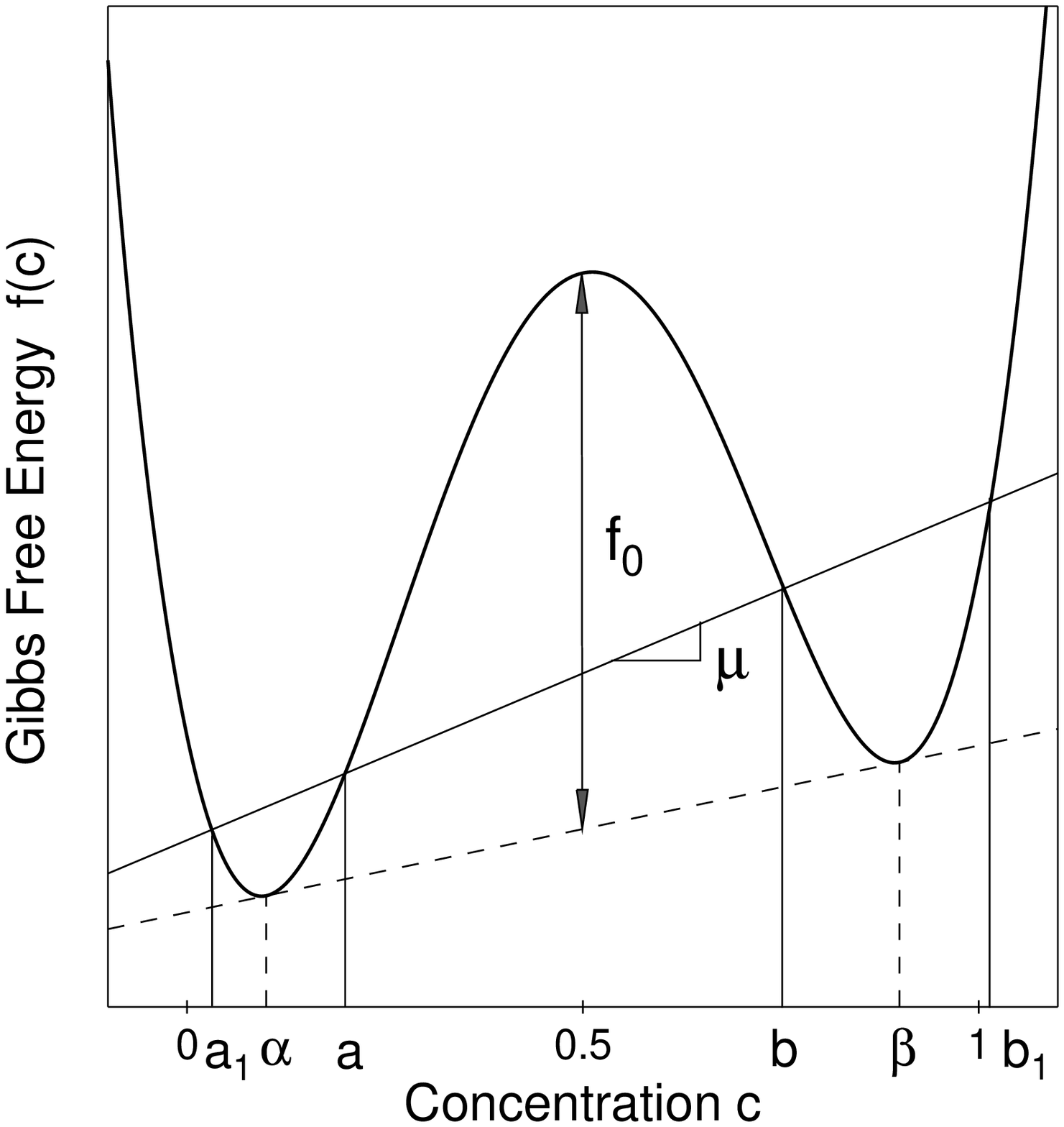}
\vspace{1cm} \hspace*{14cm} {\bf Fig. 2 } 
\newpage

\vspace*{3cm}
\epsfxsize=16cm
\epsffile{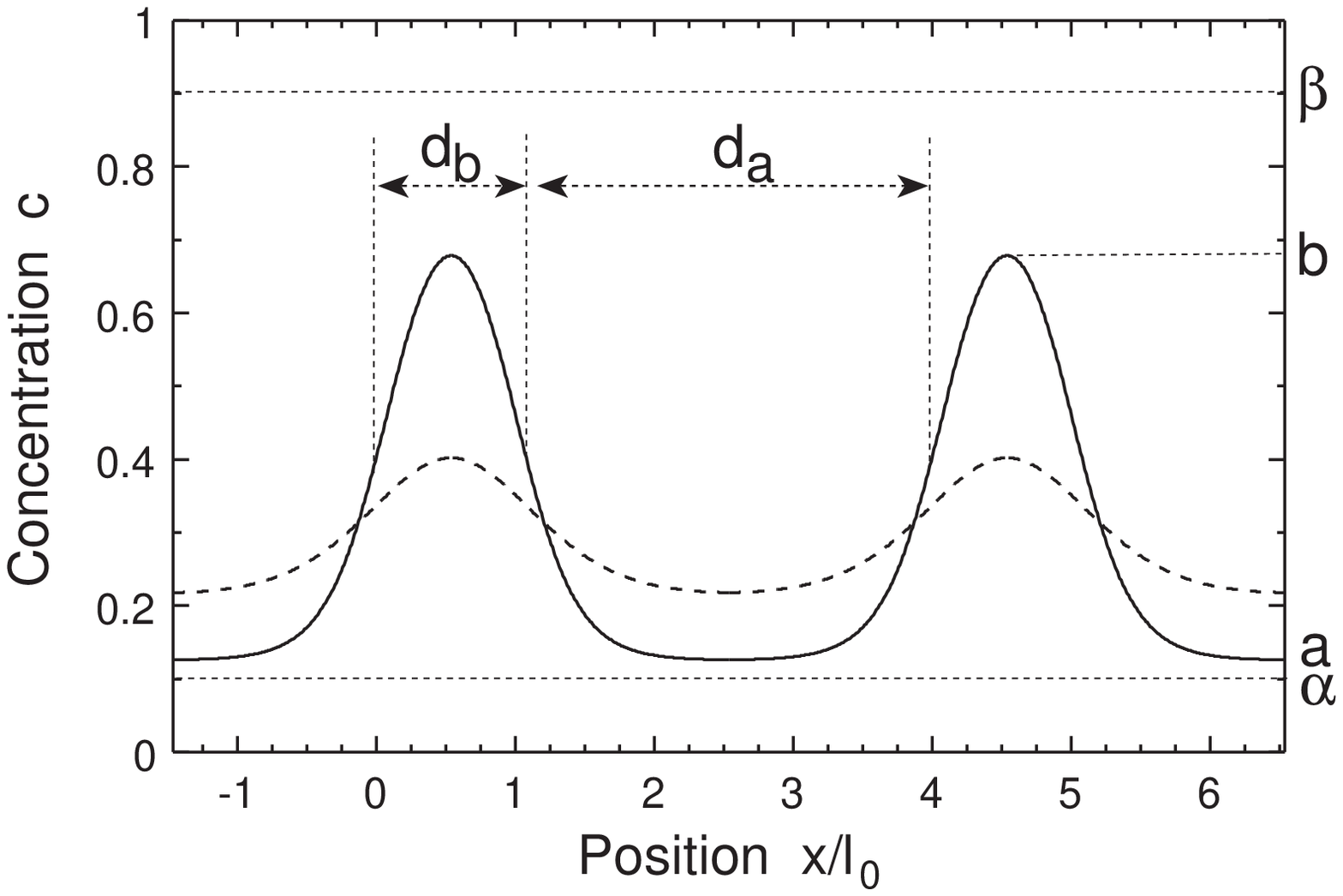}
\vspace{3cm} \hspace*{14cm} {\bf Fig. 3 } 
\newpage

\epsfxsize=16.6cm
\epsffile{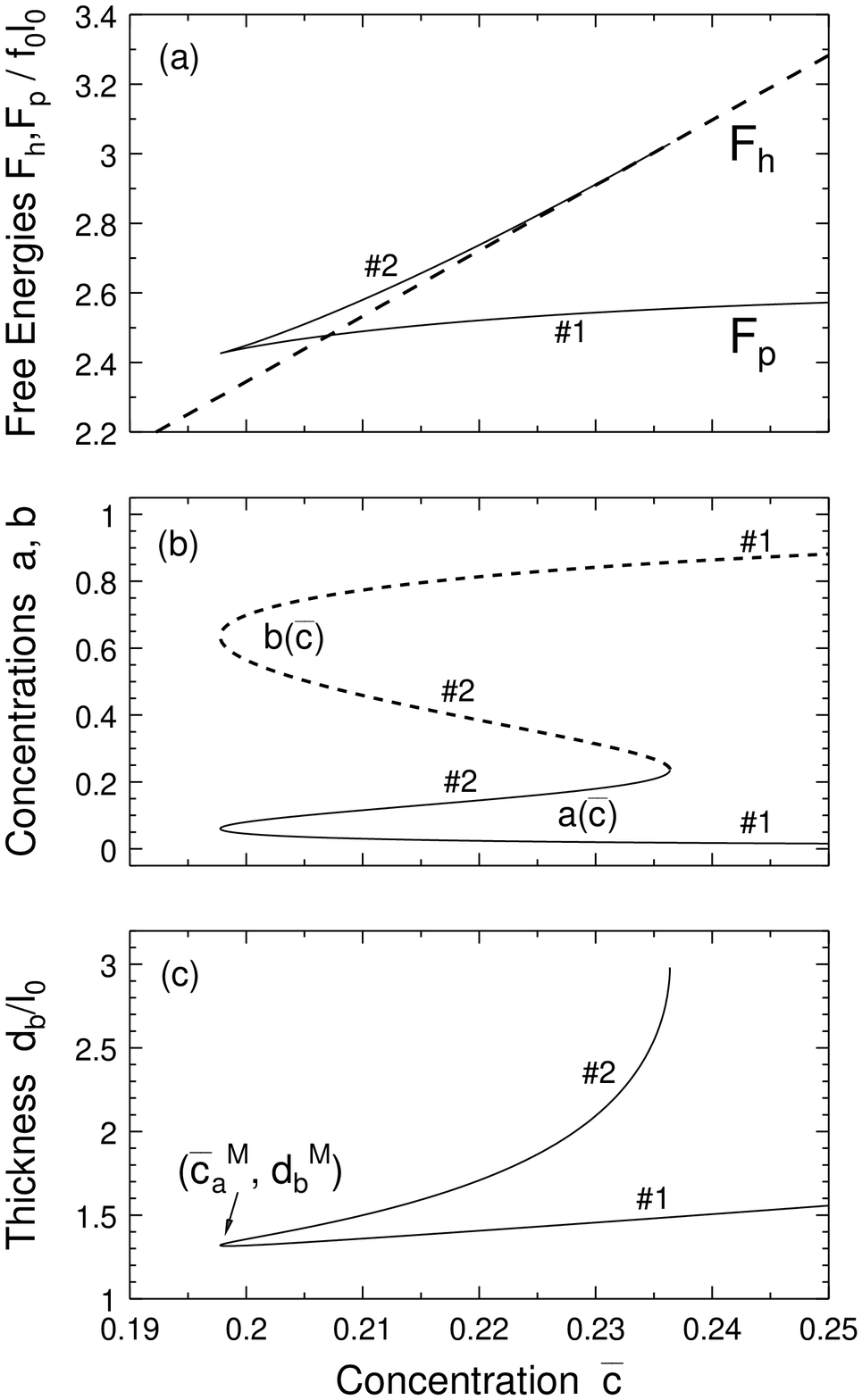}
\vspace{0cm} \hspace*{14cm} {\bf Fig. 4 }   
\newpage

\epsfxsize=16cm
\epsffile{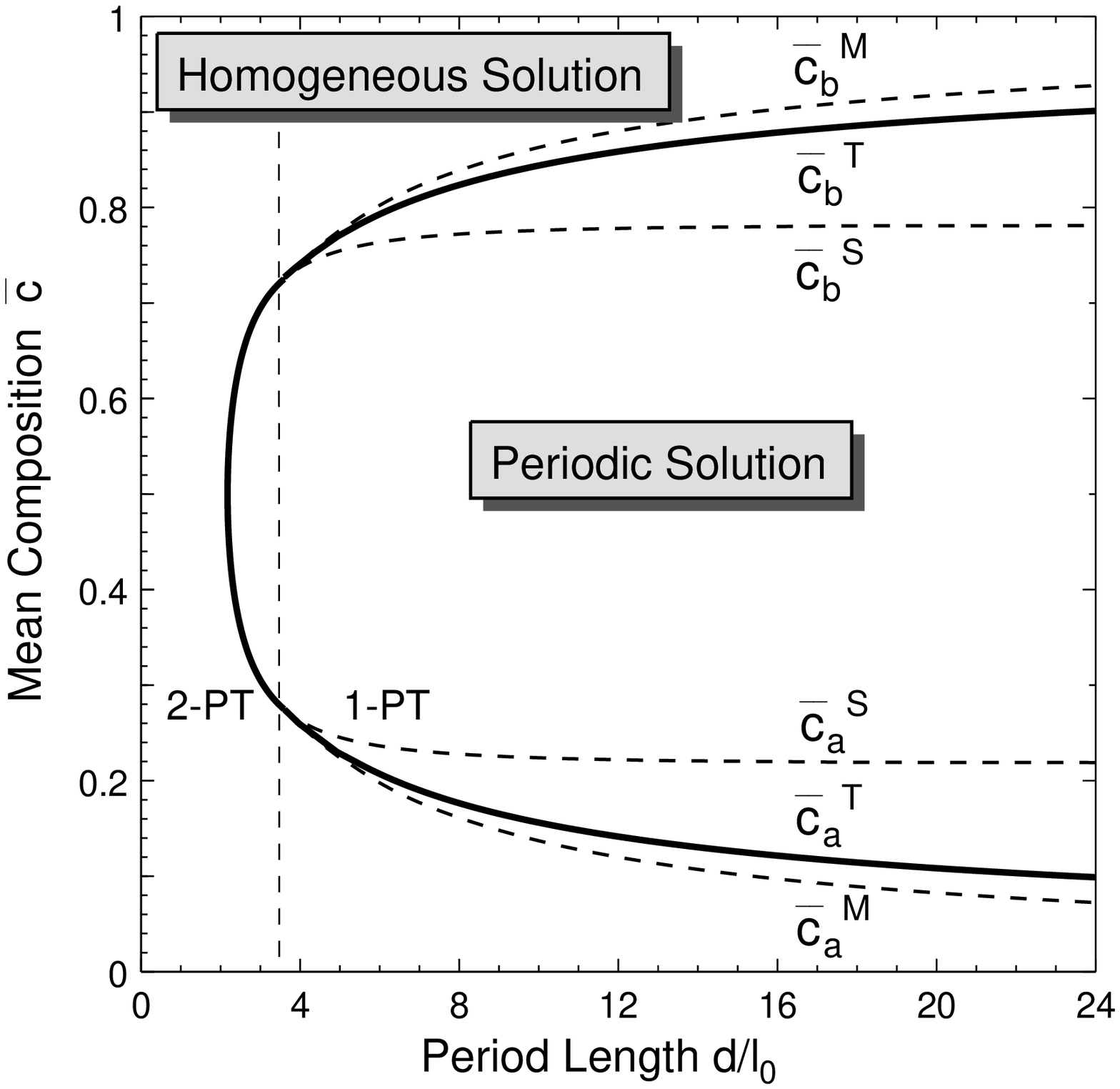}
\vspace{0cm} \hspace*{14cm} {\bf Fig. 5 }   
\newpage

\epsfxsize=14cm
\epsffile{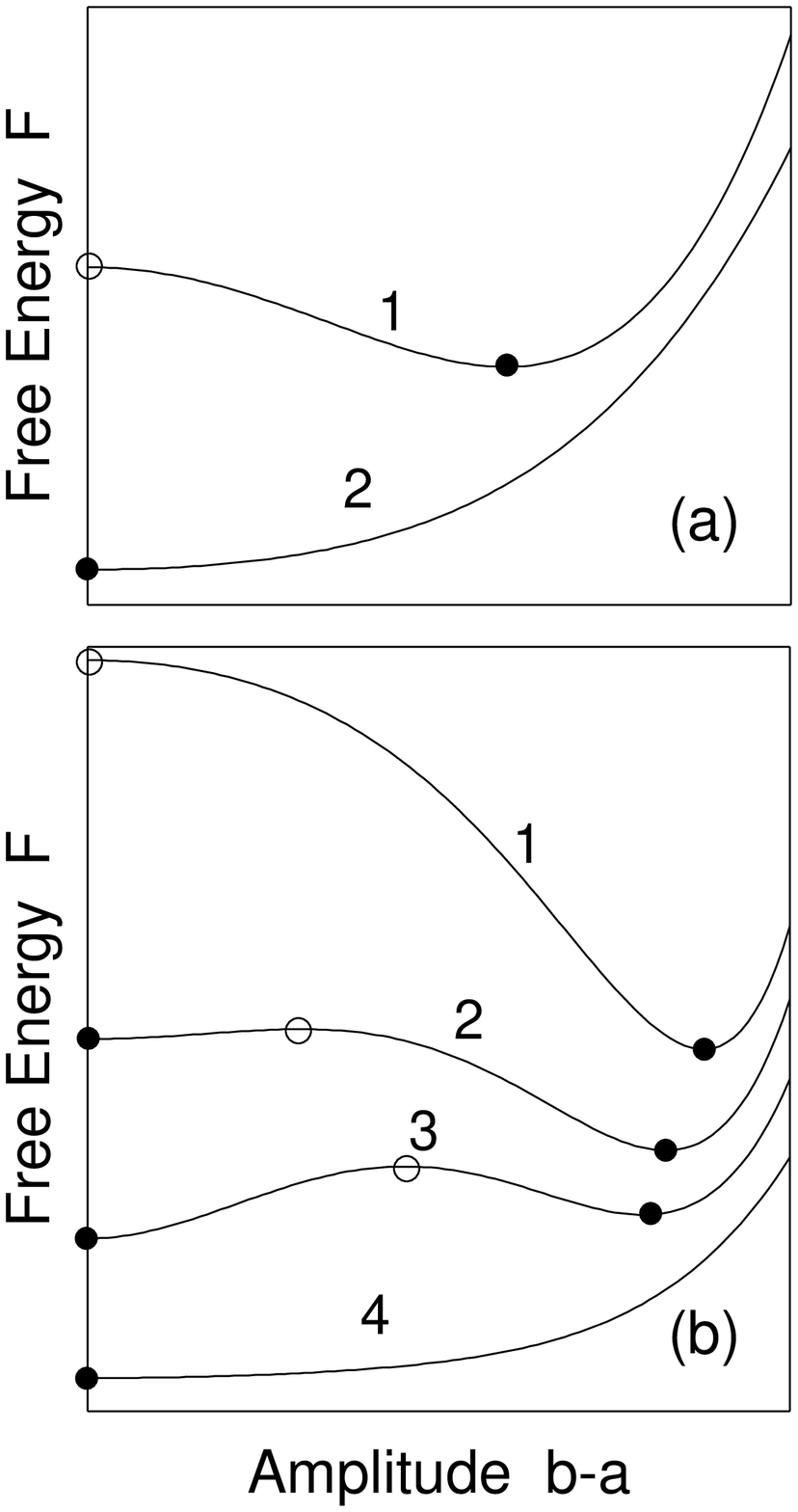}  
\vspace{0.5cm} \hspace*{14cm} {\bf Fig. 6 }
\newpage

\vspace*{3cm}
\epsfxsize=16cm
\epsffile{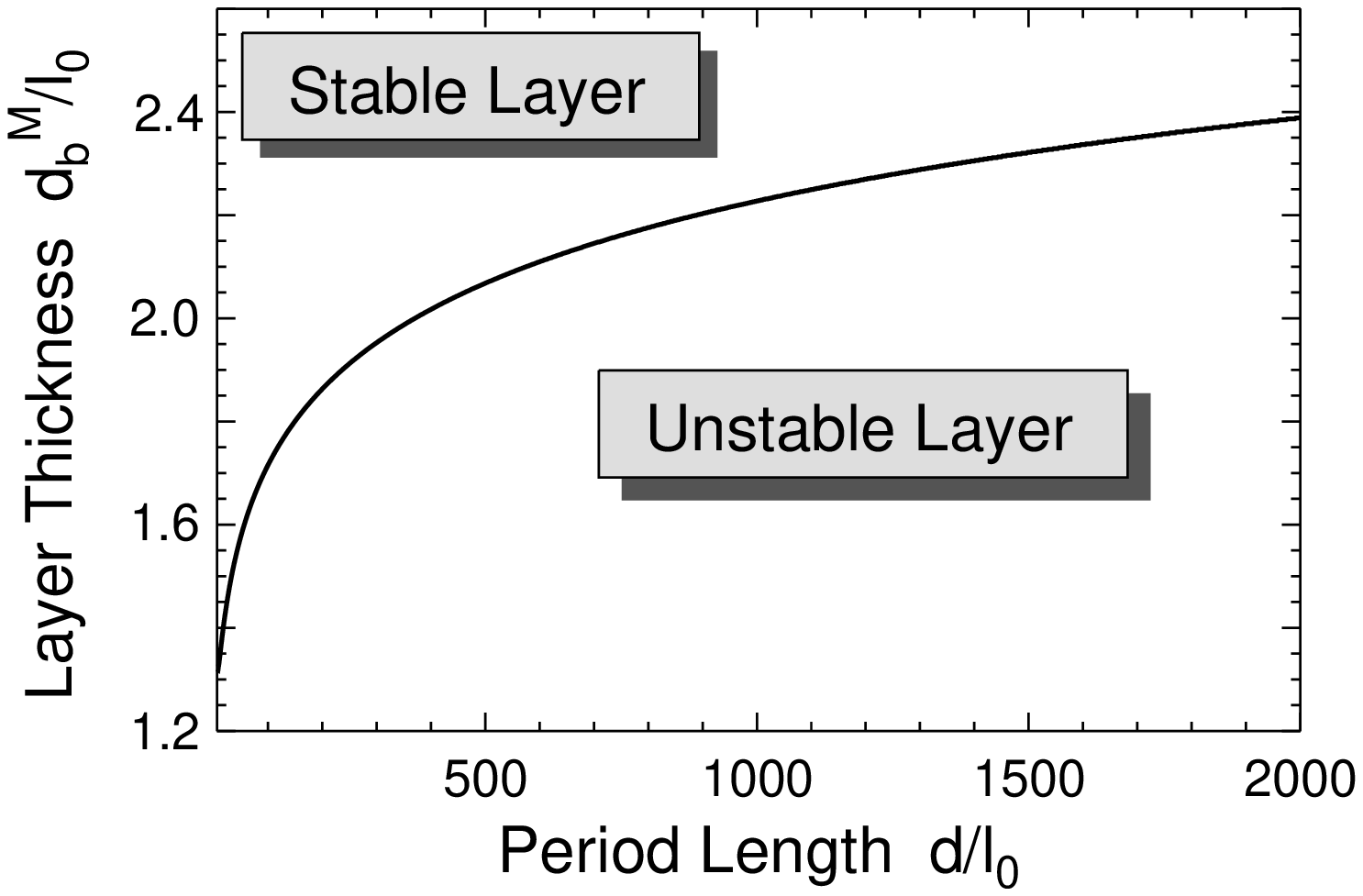}
\vspace{1cm} \hspace*{14cm} {\bf Fig. 7 }

\end{document}